\documentstyle[aas2pp4,epsf]{article}
\let\oldfootsep=\footnotesep
\def\ie{{\it {\frenchspacing i.{\thinspace}e. }}}

\def\simlt{\hbox{ \rlap{\raise 0.425ex\hbox{$<$}}\lower 0.65ex\hbox{$\sim$} }}
\def\simgt{\hbox{ \rlap{\raise 0.425ex\hbox{$>$}}\lower 0.65ex\hbox{$\sim$} }}

\def\that{{\hat t}}

\def\etal{{\it et al.}}
\def\ie{{\it i.e.}}

\def\that{{\widehat t}\,} 

\def\msun {M_\odot}
\def\mearth {M_\oplus}

\def\Amax{A_{\rm max}}

\def\ie{{\it i.e.}}
\def\lsim{\mathrel{\mathpalette\@versim<}}
\def\gsim{\mathrel{\mathpalette\@versim>}}
\def\@versim#1#2{\lower0.2ex\vbox{\baselineskip\z@skip\lineskip\z@skip
  \lineskiplimit\z@\ialign{$\m@th#1\hfil##\hfil$\crcr#2\crcr\sim\crcr}}}
\def\spose#1{\hbox to 0pt{#1\hss}}
\def\simlt{\mathrel{\spose{\lower 3pt\hbox{$\mathchar"218$}}
     \raise 2.0pt\hbox{$\mathchar"13C$}}}
\def\simgt{\mathrel{\spose{\lower 3pt\hbox{$\mathchar"218$}}
     \raise 2.0pt\hbox{$\mathchar"13E$}}}

\begin{document}

\title{Detecting Earth-Mass Planets with Gravitational Microlensing} 

\author{
	David~P.~Bennett\altaffilmark{1,2,3} and
	Sun~Hong~Rhie\altaffilmark{1}
	}

\vspace{-10mm}
\begin{abstract} 
\rightskip = 0.0in plus 1em
We show that Earth mass planets orbiting stars in the Galactic disk and
bulge can be detected by monitoring microlensed stars in the Galactic bulge.
The star and its planet act as a binary lens which generates a lightcurve which
can differ substantially from the lightcurve due only to the star itself.
We show that the planetary signal remains detectable for planetary masses
as small as an Earth mass when realistic source star sizes are included
in the lightcurve calculation. These planets are detectable
if they reside in the ``lensing zone" which is centered between 1 and 4 AU
from the lensing star and spans about a factor of 2 in distance.  If 
we require a minimum deviation of 4\% from the standard point-lens microlensing
lightcurve, then we find that more than 2\% of all $\mearth$ planets and
10\% of all $10\mearth$ in the lensing zone can be detected. If a third of
all lenses have no planets, a third have $1\mearth$ planets and
the remaining third have $10\mearth$ planets then we estimate 
that an aggressive ground based microlensing planet search program could
find one earth mass planet and half a dozen $10\mearth$ planets per year.

\end{abstract}
\vspace{-10mm}


\altaffiltext{1}{Lawrence Livermore National Laboratory, Livermore, CA 94550\\
	Email: {\tt bennett, sunhong@igpp.llnl.gov}}

\altaffiltext{2}{Center for Particle Astrophysics,
	University of California, Berkeley, CA 94720}

\altaffiltext{3}{Department of Physics, University of California,
Davis, CA 95616 }

\setlength{\footnotesep}{\oldfootsep}

\section{Introduction}
\label{sec-intro}
The recent discovery of several giant planets
(\cite{swiss51peg}, \cite{marbut}) has confirmed the existence of
planets orbiting main sequence stars other than the Sun. 
Two of these first 3 giant planets have orbits that were unexpected, and
this together with the surprising discovery of planets in a pulsar
system (\cite{wol}) demonstrates the importance of observational
studies of extra-solar planetary systems. Indirect ground based techniques
which detect the reflex motion of the parent star through accurate
radial velocity measurements or astrometry are likely to have sensitivity
that extends down to the mass of Saturn ($\sim 100\mearth$) (\cite{butetal})
or even down to $10\mearth$ (\cite{shao}) with interferometry from Keck
or the VLT.
There is great interest in searching for planets with a masses similar to
that of the Earth, and NASA's new ExNPS program (\cite{ExNPS}) seeks to build a
spacecraft capable of imaging nearby Earth mass planets in the infrared.
In order to ensure the success of such
a mission, we will need to have at least a rough idea of how prevalent
planets with masses close to that of the Earth really are.

A ground based gravitational microlensing survey system sensitive to 
planets down to $1\,\mearth$ has been proposed by Tytler (1995). 
This project
would involve both a microlensing survey telescope to detect microlensing
events in progress and a world-wide network of follow-up telescopes that
would monitor the microlensing lightcurves on a $\sim\,$hourly timescale 
in search of deviations due to planets. Existing 
microlensing surveys (\cite{macho-nat}, \cite{eros-nat}, \cite{ogle1}, 
and \cite{duo}) have recently demonstrated real time microlensing detection
capability (\cite{macho-alert}, \cite{ogle-ews}), and two
world-wide microlensing follow-up collaborations 
(\cite{planet} and \cite{macho-gman}) are now in operation, but 
to detect Earth mass planets, more capable survey and follow-up systems
will be required.

In this paper, we provide the theoretical basis for this enterprise by
calculating realistic microlensing lightcurves and detection probabilities
for planets as small as $1\,\mearth$. Previous authors
(\cite{mao-pac}, \cite{gould-loeb}, and \cite{bolatto}) have considered
the deviations from the single lens lightcurve due to planets
using the point source approximation. This is a poor approximation 
for planets in the 1-10$\,\mearth$ mass range, so we
have calculated planetary-binary lensing event lightcurves for realistic
finite size source stars, and we show that planets in
the 1-10$\,\mearth$ mass range can cause deviations from the standard single
lens lightcurve with amplitudes larger than 10\% which last for a couple
hours or more. We calculate planetary detection probabilities based upon 
a set of assumed event detection criteria and a simple planetary system
model loosely based upon the solar system.

\section{Microlensing}
\label{sec-micro}

The only observable feature of a microlensing event is the time variation
of the total magnification of all the lens images due to the motion of
the lens with respect to the observer and source.
The characteristic transverse scale
for a lens of mass $M$ is given by the Einstein ring radius which is the
radius of the ring image obtained when the source, lens and observer are
collinear. It is given by
\begin{equation} 
\label{eq-ERdef}
R_E = 2\sqrt{GMD\over c^2} = 4.03\,AU \sqrt{\left({M\over\msun}\right)
 \left({D\over 2\,{\rm kpc}}\right) },
\end{equation} 
where $D$ is the ``reduced distance" defined by $1/D = 1/D_{ol} + 1/D_{ls}$.
$D_{ol}$ and $D_{ls}$ are the distances from the observer to the lens and 
from the lens to the source respectively. 
For a point mass lens, the 
amplification of a microlensing event is given by
\begin{equation} 
\label{eq-ptAmp}
A = {u^2+2 \over u\sqrt{u^2+4}} \ ; 
    \quad u = \sqrt{u_{\rm min}^2 + (2(t-t_0)/\that)^2} ,
\end{equation} 
where $u$ is the separation
of the lens from the source-observer line of sight in units of $R_E$, and
$t_0$ and $\that$ refer to the time of peak amplification and the Einstein
diameter crossing time respectively.

In a ``planetary lensing event\rlap," the majority of the lightcurve is
described by eq.~(\ref{eq-ptAmp}), but in the region of the planetary deviation
we must consider the binary lens case (\cite{schneid}, \cite{rhie}).
If $\omega$ and $z$ denote (in complex coordinates) the source and image
positions in the lens plane, the binary lens equation is given by
\begin{equation}
\label{eq-bilens}
    \omega  = z - {1-\epsilon\over \bar z - \bar x_s} 
                - {\epsilon\over \bar z - \bar x_p} \ ,
\end{equation}
where $\epsilon$ is the fractional mass of the planet, and $x_s$ and $x_p$
are the positions of the star and planet respectively. We work in units
of the Einstein radius, $R_E$, of the total mass $M$. Eq.~(\ref{eq-bilens})
has 3 or 5 solutions ($z$) for a given source location, $\omega$.

The Jacobian determinant of the lens mapping (\ref{eq-bilens})  is
\begin{equation}
\label{eq-Jdef}
  J  = 1 - |\partial_{z}\bar\omega|^2 \quad ; \quad 
  \partial_{z}\bar\omega 
         = {1-\epsilon\over (z-x_s)^2} + {\epsilon\over (z-x_p)^2} \ ,
\end{equation}
and  the total amplification of a point source is obtained by summing up 
the absolute value of the inverse Jacobian determinant calculated at each
image:
\begin{equation}
\label{eq-Asum}
A = \sum_i |J_i|^{-1} \ . 
\end{equation}

The curve defined by $J=0$ is known as the critical curve, and the lens
mapping (\ref{eq-bilens}) transforms the critical curve to the caustic
curve in the source plane. By eq.~(\ref{eq-Asum}), a point source which
lies on a caustic will have an infinite magnification.
(The singularity at $J=0$ is integrable,
so finite sources always have finite magnifications.)
When the source star is in the
region of the caustic curve, the magnification will differ
noticeably from the single lens case allowing a ``planetary" signal
to be detected. If the planet mass is of order 1-10$\,\mearth$, the 
``caustic region"  is comparable to the size of the source star, and
the point source approximation is not appropriate.

\section{Planetary Lightcurves}
\label{sec-lc}

Because lensing conserves surface brightness, the magnification of an image
is just the ratio of the image area to the source area 
(which is given by eq.~(\ref{eq-Jdef}) for a point source). For a finite
size source, we
calculate the lens magnification in the image plane where it is
given by the sum of the image area weighted by the limb darkened
source profile assumed to have the form: $I(\theta)/I(0)=1-0.6(1-\cos\theta)$.
This avoids the magnification singularities on the caustics in the source 
plane. We integrate over the images as follows:
First, we determine the location of the ``center" of each image which
is usually the image of the center of the star. If a portion of the stellar
disk is inside a caustic when the center is outside, then we must also include
an additional double image of the included portion of the star.
We then cycle over the included images and build a numerical integration
grid centered on each image. These grids are expanded until all the grid
boundary points are outside of the images. Sometimes these image grids
can include more than one image, and in these cases, the redundant grids
are dropped from the calculation.

\begin{figure}
\plottwo{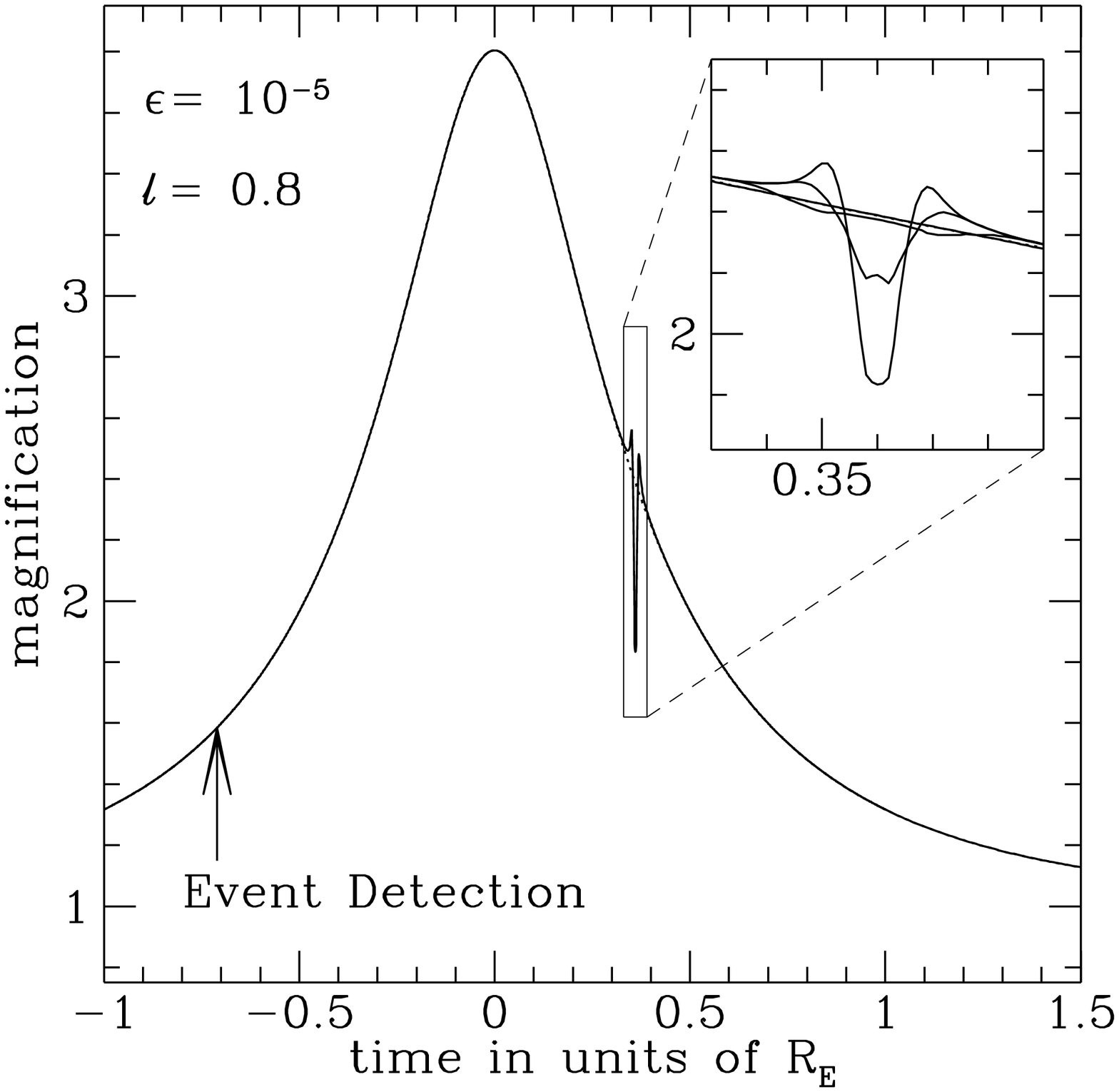}{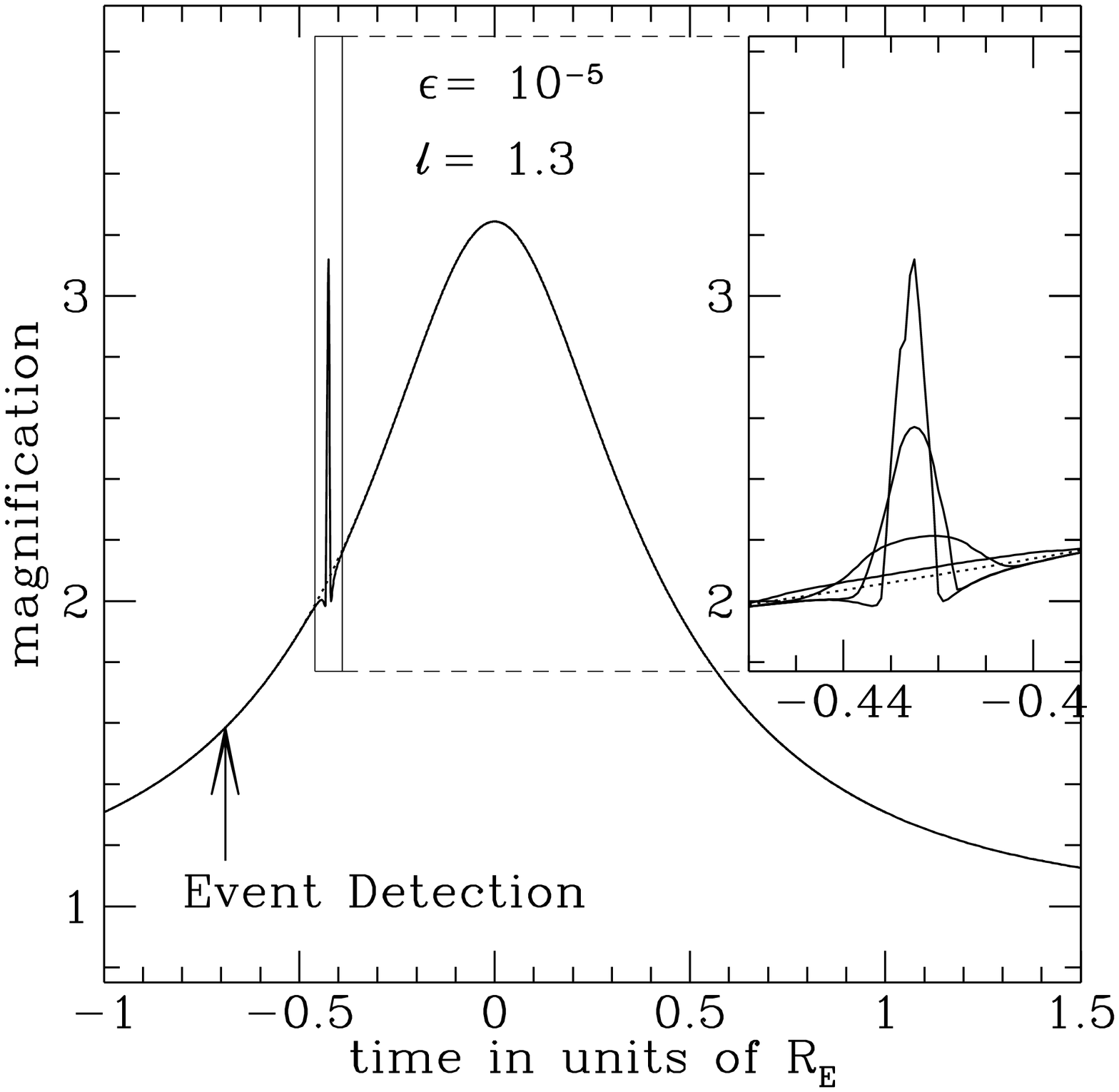}
\caption{Microlensing lightcurves which show planetary deviations are
  plotted for mass ratios of $\epsilon = 10^{-4}\ \&\ 10^{-5}$ and separations
  of $\ell = 0.8\ \&\ 1.3$. The main plots are for a stellar radius of
  $r_s = 0.003$ while the insets show light curves for radii of 0.006, 0.013,
  and 0.03 as well. The dashed curves are the unperturbed single lens
  lightcurves, $A_0(t)$. For each of these lightcurves, the source trajectory
  is at an angle of $\sin^{-1} 0.6$ with respect to the star-planet
  axis. The impact parameter $u_{\rm min} = 0.27$ for the $\ell=0.8$ plots
  and $u_{\rm min} = 0.32$ for the $\ell=1.3$ plots.
  \label{fig-lcs} } 
\end{figure}

The source radius projected to the lens plane is given by
$r_s \equiv R_s D_{ol}/(R_E (D_{ol}+D_{ls}))$ normalized to the Einstein 
ring radius.
Figure~\ref{fig-lcs} shows some examples of planetary lightcurves 
calculated for source size $r_s = 0.003$, planetary mass fraction 
$\epsilon = 10^{-5}\ \&\ 10^{-4}$, and separations of $\ell\equiv
| x_p - x_s| = 0.8\ \&\ 1.3$.
For a typical Galactic lens of $0.3 \msun$, the mass
fractions $\epsilon = 10^{-5}\ \&\ 10^{-4}$ correspond to planet masses
of $1\ \&\ 10\,\mearth$ respectively. The insets show the effects of
varying the source size $r_s$ over the values 0.003, 0.006, 0.013,
and 0.03. $r_s$ values of 0.003 and 0.006 correspond to a main sequence 
turn-off source star lensed by lenses in the disk and bulge respectively
while values of 0.013 and 0.03 correspond to a clump giant source lensed
by lenses in the disk and bulge. (We have assumed $R_s = 3R_\odot$ for
turn-off stars and $R_s = 13R_\odot$ for clump giants.) 

Figure~\ref{plate} 
shows two dimensional plots of the magnification ratio ($A/A_0$)
of the planetary binary lens case to the single lens case for the
planetary parameters $\epsilon = 10^{-4}$, $r_s = 0.003$, and
$\ell = 0.8\ \&\ 1.3$. Source trajectories are represented by
straight lines across these figures,
and the $\epsilon = 10^{-4}$ lightcurves shown in
Figure~\ref{fig-lcs} correspond to source paths which cross close to the
center of these 2-d plots at an angle of $\sin^{-1} 0.6 = 36.9^\circ$
from the horizontal lens axis.

\section{Planetary Detection Probabilities}
\label{sec-prob}

\begin{figure}
\plotone{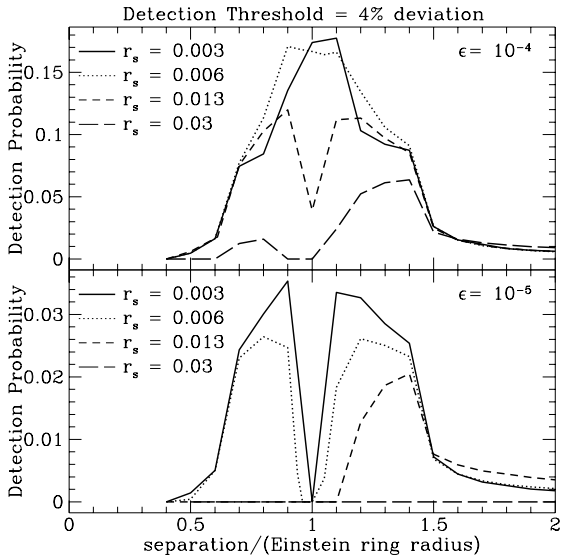}
\caption{The planetary deviation detection probability is plotted for different
  values of the planetary mass ratio, $\epsilon$, and the stellar radii,
  $r_s$. A planet is considered to be ``detected" if the lightcurve deviates
  from the standard point lens lightcurve by more than 4\% for a duration of
  more than $\that/400$. Only the portion of the lightcurve after the
  alert trigger at $A=1.58$ is searched for planetary deviations.
  \label{fig-prob} } 
\end{figure}

Let us define a reasonable set of planetary detection criteria:
First of all, the microlensing event must
be discovered by the microlensing survey system, and then the planetary
deviation must be detected by the microlensing follow-up system. The
follow-up system is assumed to observe each lensed star about
once per hour with an accuracy of 0.5-1\%
so that moderate amplitude deviations can be detected. 
Then, we require that the lightcurve deviate from the single
lens lightcurve by more than 4\% for a period longer than $\that/400$ which
is about 2.4 hours for a typical event lasting $\that = 40$ days.
This deviation must occur after the event has been detected by the survey
system which we take to be after magnification $A=1.58$ has occurred.
(This is the 0.5 magnitude event detection threshold.)
Using these detection criteria,
we examine all detectable (\ie $\Amax > 1.58$) events for a fixed
lens star-planet separation, $\ell$, and determine what fraction of these
events pass the selection criteria. This involves examining all possible
lines (or lightcurves) on the 2-d deviation plots like the ones shown in
figure~\ref{plate}. (These simulated lightcurves are assigned
uniform distributions in $u_{\rm min}$ and angle.)
Our probability results for the 4\% threshold are
shown in Figure~~\ref{fig-prob} for
fractional masses of $\epsilon = 10^{-4}\ \&\ 10^{-5}$ and a variety
stellar radii. We can see from figure~\ref{fig-prob} that
the detection probability is highest for separations close to the Einstein
ring radius, but for $\ell=1$ detection is difficult if $r_s$ is large or if
$\epsilon$ is small. A similar effect also occurs for $\ell < 1$ and
moderately large $r_s$. This can be understood by considering the
amplification contours in the source plane. 
From Figure~\ref{plate}(a)
we can see that the $\ell < 1$ lens system has a region of negative deviation
in the center of two regions containing the caustics which have a positive
deviation. A large source star will cover much of this region so that the
positive and negative deviations will tend to cancel in the integral
over the entire source. For $\ell=1$, the positive and  negative deviation
regions are more closely packed together and this effect is even stronger.

\section{A Model Planetary System}
\label{sec-model}

In order to translate the results displayed in figure~\ref{fig-prob} to
the probability of detecting planets, we must make some assumptions about
the planetary systems that we are searching for. For simplicity, let us
define a simple ``factor-of-2" model planetary system which has a distribution
of star-planet separations that is uniform in $\log(\ell)$ and has on 
average one planet for every factor of two in separation from the parent star
in the region of interest. The region of interest, or
the ``lensing zone\rlap," is the interval in $\ell$ where the
lensing detection probability is high: $0.6 < \ell < 1.5$.
For stellar lenses disk or bulge in the mass
range 0.1-$1\,\msun$, the lensing zone will cover about a factor of 2
in transverse distance somewhere in the range 0.6-6.0 AU. 
A virtue of the ``factor-of-2" model is that the distribution of planets
is not changed by orbital inclination or phase. For a planetary system
like our own where the planets' semi-major axes fall in the range
0.4-40 AU, it would be very rare for the orbital parameters to conspire to
move the outer planets inside the lensing zone, so our assumption that the
planetary system always extends through the lensing zone is reasonable.
We have used the ``factor-of-2" model to calculate the
probability of planet detection for a variety of stellar radii and detection
thresholds assuming that the planets have a unique mass.
The results are summarized in Table~\ref{tab-prob}.

\renewcommand{\thefootnote}{\fnsymbol{footnote}}

Table~\ref{tab-prob} can be used to estimate the number of planets that
might be detected with a second generation microlensing survey and follow-up
system similar to that discussed by Tytler (1995). Such a system might 
be able to discover 200 lensed turn-off stars and 50 lensed clump giants per
year\rlap.\footnote[2]{These estimates are based upon 
an assumed 2-m survey telescope which could monitor 30 million bulge stars
over a 250 day bulge season with an assumed lensing rate of 
$\Gamma = 2.4\times 10^{-5} {\rm events/yr}$ and a 50\% detection efficiency.}
We'll assume that practical difficulties such as weather serve to
reduce the actual detection probability to 50\% of the theoretical
values shown in Table~\ref{tab-prob}. Then with
planetary detection thresholds of 4\% for turn-off
stars and 2\% for clump giants (which are brighter), we would expect
to detect about 19 $10\mearth$ planets or 3 $1\mearth$ planets per year if
every lens system had a ``factor-of-2" planetary system with planets of
these masses. (These numbers are roughly independent of the fraction of the
lenses in the bulge or disk.)
If a third of all lenses have no planets, a third have
$1\mearth$ planets and the remaining third have $10\mearth$ planets, then
we would expect to detect 6 $10\mearth$ planets and a single $1\mearth$
planet every year. More than half of the $10\mearth$ planetary lightcurves
and a third of the $1\mearth$ lightcurves would have deviations larger 
than 10\%.  Clearly, a null result from an eight year survey of this
magnitude would be a highly significant indication that planetary systems
like our own are rare.

\section{Discussion}
\label{sec-char}

In the previous section, we have shown that a significant number of planets
with masses down to $1\mearth$ can be detected via gravitational microlensing
if microlensing events towards the Galactic bulge are monitored $\sim$ hourly
with photometric precision of
0.5-1.0\% which is readily achievable in crowded stellar images.

We can also use the results of our probability calculations to help determine
the optimal planetary search strategy. 
For example, given a large number of
events to monitor for planetary deviations and a limited amount of observing
time, how long should we follow each event? The probabilities given in
table~\ref{tab-prob} assume that each event is followed from event detection
at $A=1.58$ until $A$ drops to $1.13$, but if we stop the follow-up 
observations when $A>1.34$, then we will only be sensitive to planetary 
deviations from planets in the interval $0.62 < \ell < 1.62$. Integration over
the curves in figure~\ref{fig-prob} indicates that this will reduce the
chance of detecting a planet by 5-10\% (for the ``factor-of-2" model), but
the total number of observations required drops by 27\%. Thus, if the
capacity of the follow-up system is saturated, it is best to concentrate
follow-up observations on events with $A>1.34$. This effect is basically
geometric: planets that are outside the lensing zone ($\ell >1.6$)
tend to give rise to ``isolated" events that aren't associated with a
stellar lensing event detected by the survey system\rlap.\footnote[3]{Isolated
planetary lensing events might be detected by microlensing surveys, but 
the detection efficiency and variable star background rejection would be
quite poor.} It is optimal to
search for planets at $\ell < 1.6$ where they would ``modulate" a
detectable stellar lensing lightcurve.

Our results also suggest
that it will be easier to detect Earth mass planets by monitoring turn-off
star lensing events than giant star events. (Gould and Welch (1995) have
shown that combined infrared and optical
observations may allow the detection of earth mass planets
in giant star lensing events, however.)

We've established that low mass planets can be detected, but we should also 
address what can we learn about each planet that is discovered through 
microlensing. Planetary lightcurve deviations would be detected in
real time so that observations can be repeated every few minutes during the
planetary deviation.
The lens parameters $\ell$ (the separation perpendicular to
the line of sight in units of $R_E$) and the mass ratio $\epsilon$ can 
generally be determined from gross features of the lightcurve. $\ell$
is easily determined (up
to a 2-fold ambiguity) from the amplification that the unperturbed 
lightcurve would have in the deviation region, and the mass ratio $\epsilon$
can be determined from the timescale of the planetary deviation.
The 2-fold ambiguity in $\ell$ is also easily resolved in most cases
by the shape of the lightcurve deviation as can
be seen in figures~\ref{fig-lcs} and \ref{plate}. For $\ell < 1$, the
the deviation region consists of positive deviation regions surrounding the
two caustic curves with a long trench of negative deviations in between. This
leads to light curves with regions of large negative perturbations
surrounded by regions of smaller positive perturbations. For $\ell > 1$, the
situation is reversed and the dominant perturbation is a central positive
one which has regions of small negative perturbations on either side of it.

Another parameter that may be measured is the angular Einstein ring radius of 
the planet itself. This comes about because the ratio of the this radius to
the angular radius of the star is the parameter which describes the finite
source effects. For planets of Earth mass, the finite source effects are
almost always important, so in principle, this parameter may be measurable
in most events. 

In summary, we have calculated realistic lightcurves for microlensing 
events where the lens star has a low mass planetary companion, and we have
shown that planets with masses as small as $1 \mearth$ can be detected
via gravitational microlensing. Thus, gravitational microlensing is
the only ground based method that has been shown to be sensitive to
Earth mass planets.

\acknowledgments
\section*{Acknowledgements}

We'd like to thank all members of the microlensing planet search study
group and, in particular, David Tytler for stimulating our work on this
subject and for many useful discussions. We'd also like to thank Stan
Peale for extensive comments on an early draft.
Work performed at LLNL is supported by the DOE under contract W7405-ENG-48.
Work performed by the Center for Particle Astrophysics on the UC campuses
is supported in part by the Office of Science and Technology Centers of
NSF under cooperative agreement AST-8809616.

\clearpage
\begin{deluxetable}{lccccc}  
\tablewidth{0pt}
\tablecaption{Planetary Detection Probabilities \label{tab-prob} }
\tablehead{
\colhead{$r_s$} & $\epsilon$ &
\colhead{$P(2\%)$} & \colhead{$P(4\%)$} &
\colhead{$P(10\%)$} & \colhead{$P(20\%)$}
}  

\startdata
0.003 & $10^{-4}$ & 0.188 & 0.144 & 0.094 & 0.052 \nl
0.006 & $10^{-4}$ & 0.238 & 0.159 & 0.085 & 0.043 \nl
0.013 & $10^{-4}$ & 0.201 & 0.118 & 0.052 & 0.014 \nl
0.03  & $10^{-4}$ & 0.120 & 0.035 & 0.012 & 0.000 \nl
0.003 & $10^{-5}$ & 0.060 & 0.034 & 0.014 & 0.004 \nl
0.006 & $10^{-5}$ & 0.052 & 0.026 & 0.005 & 0.002 \nl
0.013 & $10^{-5}$ & 0.019 & 0.008 & 0.001 & 0.000 \nl
0.03  & $10^{-5}$ & 0.002 & 0.000 & 0.000 & 0.000 \nl
\enddata
\tablenotetext{} {Planetary detection probabilities $P$ are shown as a
 function of the deviation threshold for different values of the
 source star radius $r_s$, and planetary mass fraction $\epsilon$.
 Idealized ``factor-of-2" planetary systems with one
 planet per factor of 2 in distance from the lens star are assumed.
 A planet is considered to be detected if it
 deviates from the single lens light curve by more than the threshold for
 a period of time longer than $\that/400$. The $r_s$ values of 0.003 and
 0.006 correspond to a turn-off source star with disk and bulge lenses
 respectively, while the $r_s$ values of 0.013 and 0.03 correspond to
 a giant source with disk and bulge lenses.}
\end{deluxetable} 

%
%
\clearpage
\singlespace

%

\onecolumn 




\onecolumn

\begin{figure}
\plottwo{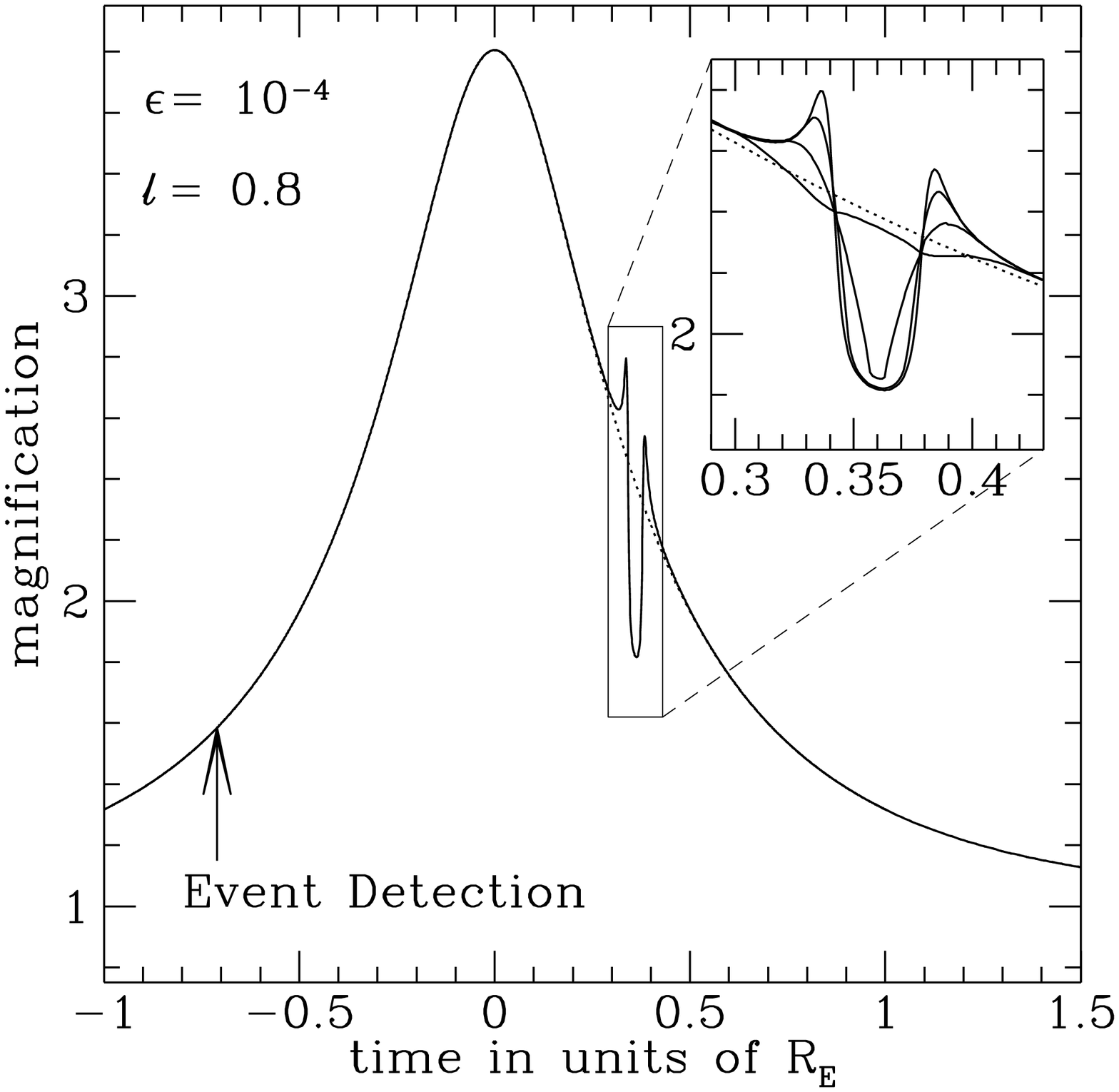}{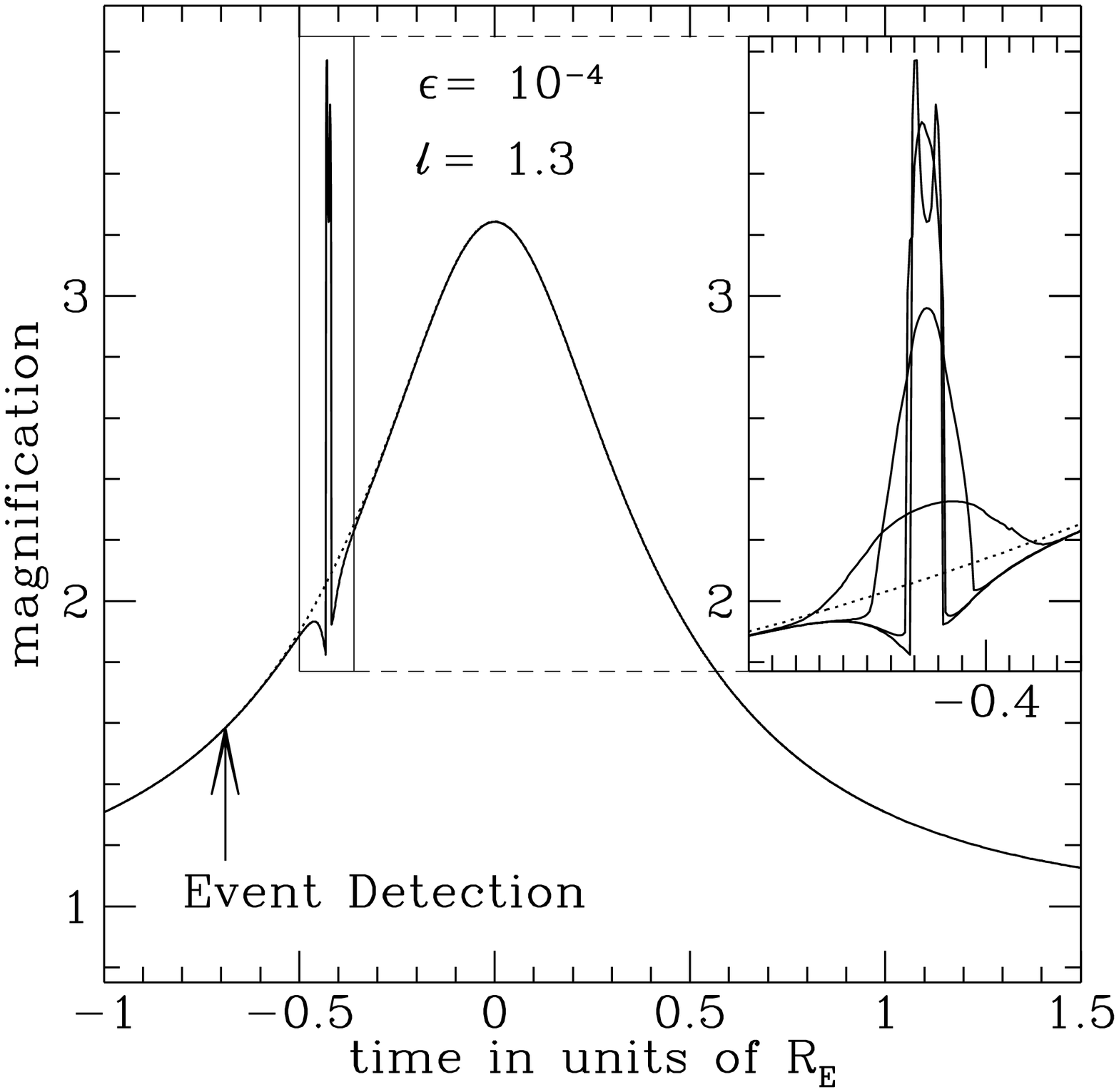}
\end{figure} 

\twocolumn



\begin{figure}
\figcaption{This plate shows the magnification ratio between the planetary
lensing case ($A$) and the single lens case ($A_0$)
as a function of source position
for $\epsilon = 10^{-4}$, $r_s = 0.003$ 
and $\ell = 0.8$ (a) \&\ $\ell = 1.3$ (b). 
\label{plate}}
\end{figure} 

\end{document}